\documentclass[final,authoryear,5p,times,twocolumn]{elsarticle}

\usepackage{amsmath}
\usepackage{amssymb}
\usepackage{lineno,hyperref}
\modulolinenumbers[5]

\usepackage{fixltx2e}
\usepackage{afterpage}

\journal{Journal of \LaTeX\ Templates}

\biboptions{authoryear}

\usepackage{color}

\begin{document}

\begin{frontmatter}

\title{Nature of the low-energy, $\gamma$-like background for the Cherenkov Telescope Array}


\author[afil]{Julian Sitarek}
\ead{jsitarek@uni.lodz.pl}
\author[afil]{Dorota Sobczy\'nska}
\ead{dsobczynska@uni.lodz.pl}
\author[afil]{Micha\l\ Szanecki}
\ead{mitsza@uni.lodz.pl}
\author[afil]{Katarzyna Adamczyk}
\ead{kadamczyk@uni.lodz.pl}
\author[afil2]{Paolo Cumani}
\ead{pcumani@ifae.es}
\author[afil2]{Abelardo Moralejo}
\ead{moralejo@ifae.es}

\address[afil]{Department of Astrophysics, The University of \L\'od\'z, ul. Pomorska 149/153, 90-236 \L\'od\'z, Poland}
\address[afil2]{Institut de Fisica d'Altes Energies (IFAE), The Barcelona Institute of Science and Technology, Campus UAB, 08193 Bellaterra (Barcelona), Spain}

\begin{abstract}
The upcoming Cherenkov Telescope Array (CTA) project is expected to provide unprecedented sensitivity in the low-energy ($\lesssim 100$\,GeV) range for Cherenkov telescopes. 
Most of the remaining background in this energy range results from misidentified hadron showers. 
In order to fully exploit the potential of the telescope systems it is worthwhile to look for ways to further improve the available analysis methods for $\gamma$/hadron separation. 
We study the composition of the background for the planned CTA-North array by identifying events composed mostly of a single electromagnetic subcascade or double subcascade from a $\pi^0$ (or another neutral meson) decay. 
We apply the standard simulation chain and state-of-the-art analysis chain of CTA to evaluate the potential of the standard analysis to reject such events.
Simulations show a dominant role of such single subcascade background for CTA up to energies $\sim70$\,GeV. 
We show that a natural way of rejection of such events stems from a shifted location of the shower maximum, and that the standard stereo reconstruction method used by CTA already exploits most of expected separation. 
\end{abstract}


\end{frontmatter}

\def\ses{\emph{SES}}
\def\spis{\mbox{\emph{S}$\pi^0$\emph{S}}}
\newcommand{\progname}[1]{{\fontfamily{pcr}\selectfont #1}}

\section{Introduction}

The imaging air Cherenkov technique has been successfully used since the first $\gamma$-ray source (Crab Nebula) was discovered by Whipple collaboration \citep{whipple}.  
The main idea of the technique is based on the measurement of Cherenkov photons produced in the atmosphere by the charged relativistic particles from an Extensive Air Shower (EAS). 
The two dimensional angular distribution of Cherenkov light appears on the telescope camera as the shower image. Due to the fact that the number of registered hadron-induced events (the so-called background) is several orders of magnitude larger than the number of registered $\gamma$-ray events from a source, the $\gamma$/hadron separation method plays a crucial role in the data analysis. 
The image parameterization that was suggested in \cite{hi85} allowed an effective $\gamma$-ray selection. 
More sophisticated selection methods are being used now (such as \citealp{kraw2006,al08,ohm09,par14}), but most of them are still based on the original Hillas parameters.
In the last 30 years, the construction of larger mirror dish telescopes and employing stereoscopic technique allowed to lower the observation energy threshold.  
Currently three large IACT (Imagining Air Cherenkov Telescopes) instruments are in operation: H.E.S.S. \citep{aha06}, MAGIC \citep{al16a} and VERITAS \citep{weekes2002,hold11}. 

The upcoming Cherenkov Telescope Array (CTA) \citep {actis11,acha13} was designed to study $\gamma$-ray sources in a broad energy range, from a few tens of~GeV to hundreds of~TeV. 
It will consist of two arrays: one in the Northern and one in the Southern hemisphere. 
The former one will be mostly focused on lower energy observations. 
CTA is expected to have an order of magnitude better sensitivity than the currently operating IACT systems \citep{be13}. 
However at low energies the $\gamma$/hadron separation becomes more difficult, which results in the deterioration of the sensitivity. 
Apart from the fact that images are smaller, which results in a worsening of the telescope performance (the shower has to be reconstructed from information in only a few pixels), the decrease of the $\gamma$/hadron separation efficiency can be explained by a several physical effects. 
First, at low energies the geomagnetic field has more impact on $\gamma$-ray showers (making them appear more hadron-like) than hadron initiated showers thus the efficiency of primary  particle selection is worse (see e.g. \citealp{bo92,comm08,sz13}). 
Second, larger fluctuations of the image parameters are expected due to the larger fluctuations of the Cherenkov light density at ground in the low energy region \citep{bhat,so09a}. 
Third, the $\gamma$-ray events may be imitated by a specific type of a hadron-induced shower. 
It has been suggested in \cite{maier} that hadronic events that survive the $\gamma$-ray selection criteria, transfer much of the primary's energy to electromagnetic sub-cascades during the first few interactions. 
Furthermore, it has been shown in \cite{sob2007} that a large telescope can be triggered by light produced by a $e^+/e^{-}$ from only one or two electromagnetic sub-cascades, which are produced by a single $\pi^0$ decay in the hadron initiated shower. 
These images have very similar shapes to $\gamma$-ray events, therefore they can be called $\gamma$-like background. 
Both single sub-cascade and single $\pi^{0}$ events are mainly proton induced showers with relatively low primary energy (below $\sim 200$\, GeV). 
The efficiency of the $\gamma$/hadron separation method based on the parameters describing the image shape, decreases at low energies due to the occurrence of this specific $\gamma$-like background events for the IACTs system \citep{so15a,sob15b}.

It should be also noted that a primary electron or positron \citep{ams14} can induce an EAS that triggers the system of IACTs \citep{co01,elhess}. 
The background from a $e^{-}/e^{+}$ is hardly distinguished from a $\gamma$ ray as both form pure electromagnetic cascades in the atmosphere. 
The cosmic ray electron spectrum is however steeper then the one of protons, resulting in a complicated energy dependence of this type of background. 

We study the impact of such events on the Northern CTA array by using Monte Carlo (MC) simulations. 
In particular, we investigate the ability of the state-of-the-art CTA analysis methods to reject such a background. 
In Section~\ref{sec:sim} we describe the performed MC simulations and the analysis methods.
In Section~\ref{sec:res} we report the obtained results on the expected background structure for CTA North array. 
We discuss and summarize the results in Section~\ref{sec:con}.

\section{Simulation pipeline}\label{sec:sim}
The simulations of atmospheric showers were performed with a modified code of \progname{CORSIKA} 7.5 \citep{he98}.
We use UrQMD \citep{ba98} and QGSJET-II-04 \citep{os11} as the low (particle energy $<80$\,GeV) and high energy hadronic models respectively. 
According to \cite{maier} hadronic interactions with low multiplicity and high $\pi^{0}$ fraction in the first stage of the shower development may result in the occurrence of hadronic images that survive the $\gamma$ selection. 
Thus the estimation of the hadronic background after the $\gamma$/hadron separation could depend on the chosen interaction model. 
However, \cite{so15a} showed that the fractions of a single $\gamma$ or $\pi^{0}$ events in protonic background obtained with different interaction models do not differ significantly for stereo systems. 
The author demonstrated that the effect of this hardly reducible background is more sensitive on the altitude of the observatory, telescope size and trigger conditions than on the interaction model.

The response of the telescopes was simulated using the \progname{sim\_telarray} code \citep{be08}.
The output was converted with \progname{Chimp} \citep{ha15,ha17} to allow analysis using MAGIC Analysis and Reconstruction Software (MARS) \citep{za13,al16b}.

\subsection{MC samples}
The EAS development was simulated, using the \progname{CORSIKA} code, for the North CTA site, located at 2147 m a.s.l. on the Canary island of La Palma.
We simulated $\approx 10^{8}$  $\gamma$ rays and an order of magnitude more hadronic background events - composed of protons and He nuclei. 
We simulated also a sample of electron background events. 
Detailed information concerning the used EAS simulations settings are shown in Table~\ref{tab:mc_cor}.

\begin{table*}[t]
\begin{center}
 \begin{tabular}{|c||c|c|c|c|}
  \hline
  \progname{CORSIKA} input & \multicolumn{4}{|c|}{Input value}\\
  \cline{1-2}
  \hline \hline
  Primary Particle type  &  $\gamma$ ray  & \multicolumn{3}{|c|}{Background components}\\
  \cline{3-5}
 & & Proton & Helium & Electron\\
  \hline
  Energy Range [GeV] & $5$ - $2000$ & $8$ - $4000$& $16$ - $8000$ &  $5$ - $2000$\\
  \hline
  Energy power-law index & \multicolumn{4}{|c|}{-2.0}\\
  \hline
  Impact Parameter [m] & 0 - 1100 & \multicolumn{3}{|c|}{0-1600}\\
  \hline
  Zenith Angle [$^{\circ}$] & \multicolumn{4}{|c|}{20}\\
  \hline
  Azimuth Angle (AZ) [$^{\circ}$] & \multicolumn{4}{|c|}{0 \&180}\\
  \hline
  Event reusage  & 5 &  \multicolumn{3}{|c|}{20}\\
  \hline
  Viewcone [$^{\circ}$]& 0 & \multicolumn{3}{|c|}{10}\\
  \hline
  Number of events &  $9.5\times10^{6}$ &  $2.4\times10^{7}$& $1.53\times10^{7}$ & $9.5\times10^{6}$ \\
 (per each AZ, not reused) & & & & \\
  \hline
   \end{tabular}
\end{center}
\caption{Main \progname{CORSIKA} steering  parameters used in the simulations.}
 \label{tab:mc_cor}
\end{table*}

\begin{figure}[t]
  \begin{center}
    \includegraphics[scale=0.4]{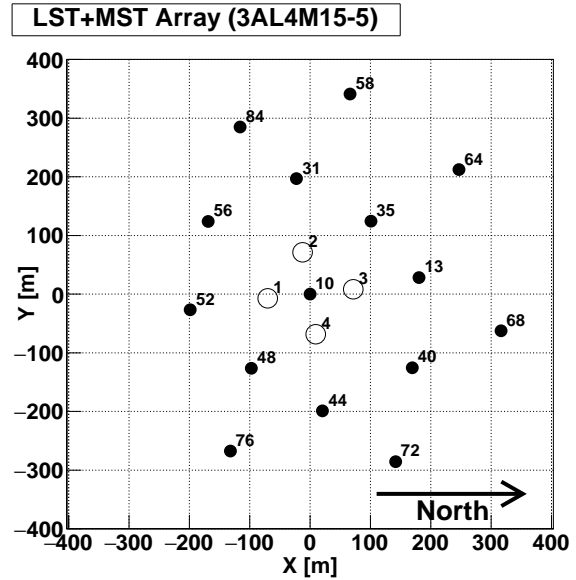}
    \caption{
Telescope array used in our simulations. 
The LST type telescopes are marked with empty circles. 
Full circles represent MST telescopes. 
The numbering of the telescopes follows the \progname{Production-3} convention. 
The geographic north direction is indicated with an arrow. }
    \label{fig:Array_tel1}
  \end{center}
\end{figure}

The geophysical parameters of the chosen La Palma site are set accordingly to the standard La Palma site configuration template of  \progname{CORSIKA} steering card used in the standard CTA simulation package - \progname{corsika\_simtelarray}. 
We use the official CTA configuration settings, the so-called \progname{Production-3}. 
The individual telescopes' position (see Fig. ~\ref{fig:Array_tel1}), takes into consideration also the orography of the La Palma site, i.e. the Z coordinate of the telescopes changes with the position to include the difference in altitude. 

The simulated arrival zenith angle (ZA) is set to $20^{\circ}$ for $\gamma$ rays, whereas for background events we use a diffuse viewcone with a half-opening angle of  $10^{\circ}$, centred at the same ZA=$20^{\circ}$, (for details see Table~\ref{tab:mc_tel}). 
As the geomagnetic field (GF) affects the detection and reconstruction performance of IACT, to recover the impact of this effect, we simulated two opposite azimuth angles of arrival of primary particles AZ=0$^{\circ}$ and AZ=180$^{\circ}$, corresponding to the largest difference in magnitude of GF  (see \citealp{sz13}). 

To simulate  the telescope response to EAS we used the standard CTA software  \progname{sim\_telarray}. 
We studied the  telescope array layout presented in Fig.~\ref{fig:Array_tel1}  with baseline parameters from \progname{Production-3}. 
The layout studied here is similar to the currently planned layout of telescopes for CTA North array. 
This array is composed of 4 Large Sized Telescopes (LST) and 15 Middle Sized Telescopes (MST) and is one of the most efficient ones studied for this site \citep{ho17}.
The most important parameters used in the simulations of both kinds of telescopes are summarized in Table~\ref{tab:mc_tel}.
\begin{table*}[t]
\begin{center}
 \begin{tabular}{|c||c|c|}
  \hline
  \progname{sim\_telarray} input & \multicolumn{2}{|c|}{Input value}\\
  \cline{1-2}
  \hline \hline
  Telescope type & LST & MST \\
  \hline
  Dish type & Parabolic & Davies Cotton\\
  \hline
  Camera Focal length & ${\rm f}=28 {\rm m}$ & ${\rm f}=16 {\rm m}$\\
  \hline
  Total projected mirror area & ${\rm D}=386.9 \;{\rm m^{2}}$ & ${\rm D}=103.9 \;{\rm m^{2}}$\\
  \hline
  Camera  Field of view &  ${\rm FoV}=4.3^{\circ}$ &  ${\rm FoV}=7.7^{\circ}$\\
  \hline
  Pixel size & $0.1^{\circ}$ & $0.18^{\circ}$\\

  \hline
  Number of Pixels & \multicolumn{2}{|c|}{$1855$}\\
  \hline
  Photomultipliers type/ & Hamamatsu R11920 &Hamamatsu R12992\\
  quantum efficiency &  ${\rm QE}_{\rm peak}=40.8\%$ & ${\rm QE}_{\rm peak}=43.2\%$\\
  \hline
  \hline
  Telescope trigger type & \multicolumn{2}{|c|}{Analog sum} \\
  \hline
Trigger threshold & Sum of $\approx$43 phe & Sum of $\approx$46 phe\\
(amplitude) &  in 21 pixels & in 21 pixels\\
  \hline
  Average NSB (per telescope) & 0.317 phe/ns & 0.276 phe/ns\\
 \hline
  Min.\ trigger multiplicity & \multicolumn{2}{|c|}{2 telescopes of the same kind} \\
\hline
   \end{tabular}
\end{center}
\caption{Parameters assumed in \progname{sim\_telarray} for our simulations.}
 \label{tab:mc_tel}
\end{table*}
As our studies focus mainly on the lower energies, we also separately investigate the LST subarray (telescopes 1-2-3-4 in Fig.~\ref{fig:Array_tel1}). 
All the analysis steps are done individually for both arrays. 

\subsection{Analysis}\label{sec:anal}
During the simulations of an air shower we mark the occurrence of a Single Electromagnetic Subcascade, hereafter \ses\ and Single $\pi^0$ Subcascades, hereafter \spis .
We define \ses\ as a particle (normally e$^\pm$ or a $\gamma$ ray) undergoing an electromagnetic interaction, and all the secondary particles created in the electromagnetic cascade starting from that particle. 
Similarly we define \spis\ as the primary particles created in a decay of a neutral particle (normally $\pi^0$ or $\eta$) and all the secondary particles created in the subshower started by these particles. 
Each new \ses\ and \spis\ generated in the shower have a unique number assigned to it. 
Each Cherenkov photon produced in the shower have two additional numbers propagated, to identify its corresponding \ses\ and \spis , unless it was created by a muon or by a charged hadron (e.g. the primary particle). 

For the Cherenkov photons that are reflected from the telescope mirror dish and are converted into photoelectrons (phe) in the camera we calculate the statistics of \ses\ and \spis .
A \ses\  (or \spis ) is considered to contribute to the event if it produced at least 6\,phe in at least one of the triggered telescopes. 
The value was selected to be similar to the cleaning level applied later on in the analysis. 
From the \ses\  and \spis\ that satisfy the above condition  we calculate the numbers of \ses\  and \spis\ participating in a given event.
For each \ses\ we compute also a ratio of a number of phe originating from it to the total number of phe measured in all the triggered telescopes. 
We call \ses$_{\max}$ the largest of these ratios (i.e. for the most dominating \ses ), and similarly \spis$_{\max}$ for the most dominating \spis .
We call an event \ses -dominated if \ses$_{\max}>70\%$ and  \spis -dominated if \spis$_{\max}>70\%$. 
According to this definition the same event can be both \ses -dominated and \spis -dominated, and in fact if the largest \ses\ in a \ses -dominated event comes from a decay of e.g. $\pi^0$ it will automatically make it a \spis -dominated event as well.
On the other hand, an event composed of a similar amount of light registered from two separate \ses\ will not qualify as \ses -dominated as in this case \ses$_{\max}\approx 50\%$.
A \ses -dominated event that is not \spis -dominated event is as well possible (however not very common). 
An example of such a process is a $\pi^\pm$ produced high in the atmosphere decaying to $\mu^\pm$ which in turn decays to $e^\pm$. 
The information about dominating \ses\ and \spis\ and about the total number of \ses s and \spis s that contribute to the event is then propagated through the analysis chain. 

\begin{figure}[pt!]
  \begin{center}
    \includegraphics[width=0.45\textwidth]{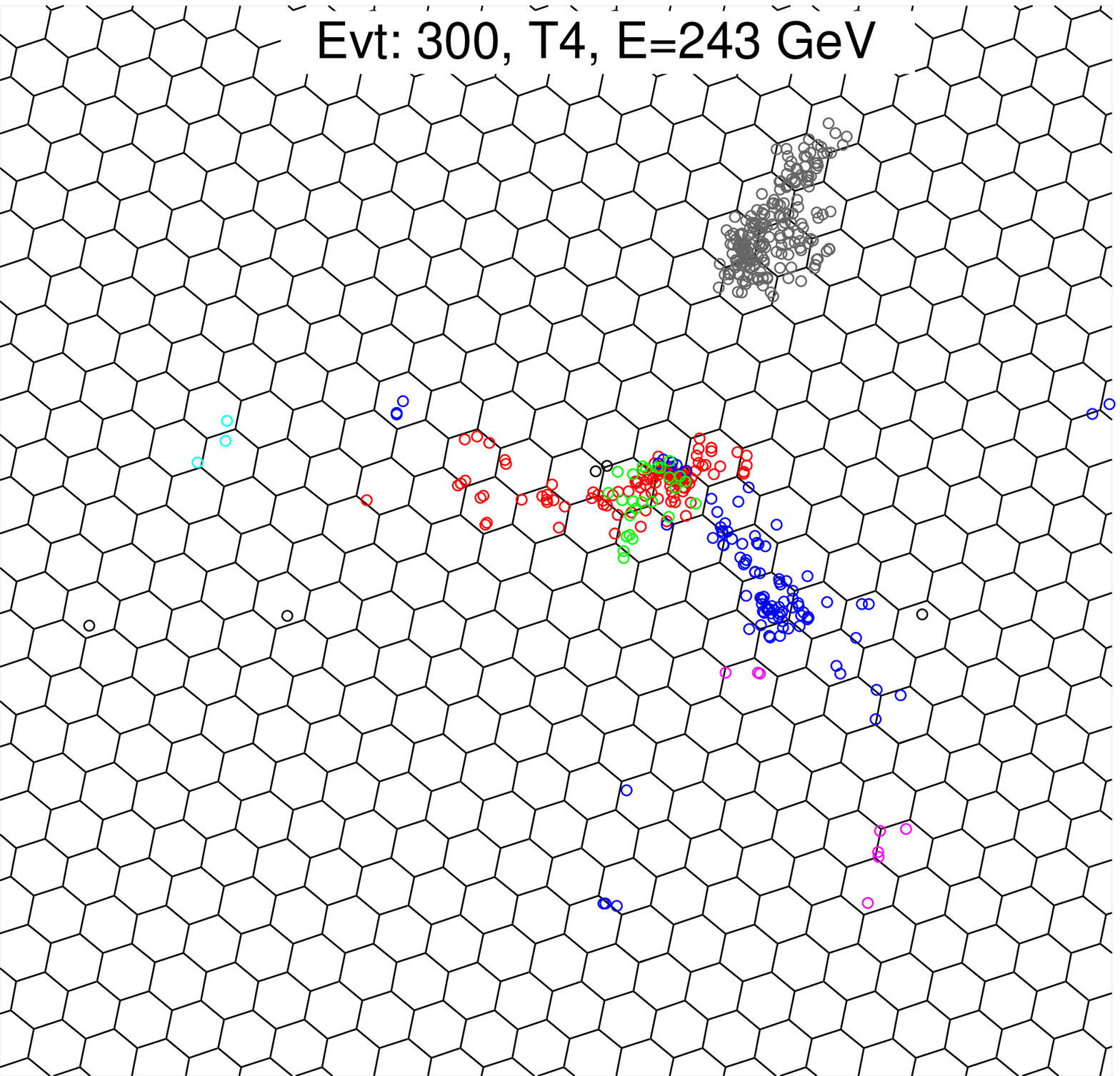}\\ \vspace {0.5cm} 
    \includegraphics[width=0.45\textwidth]{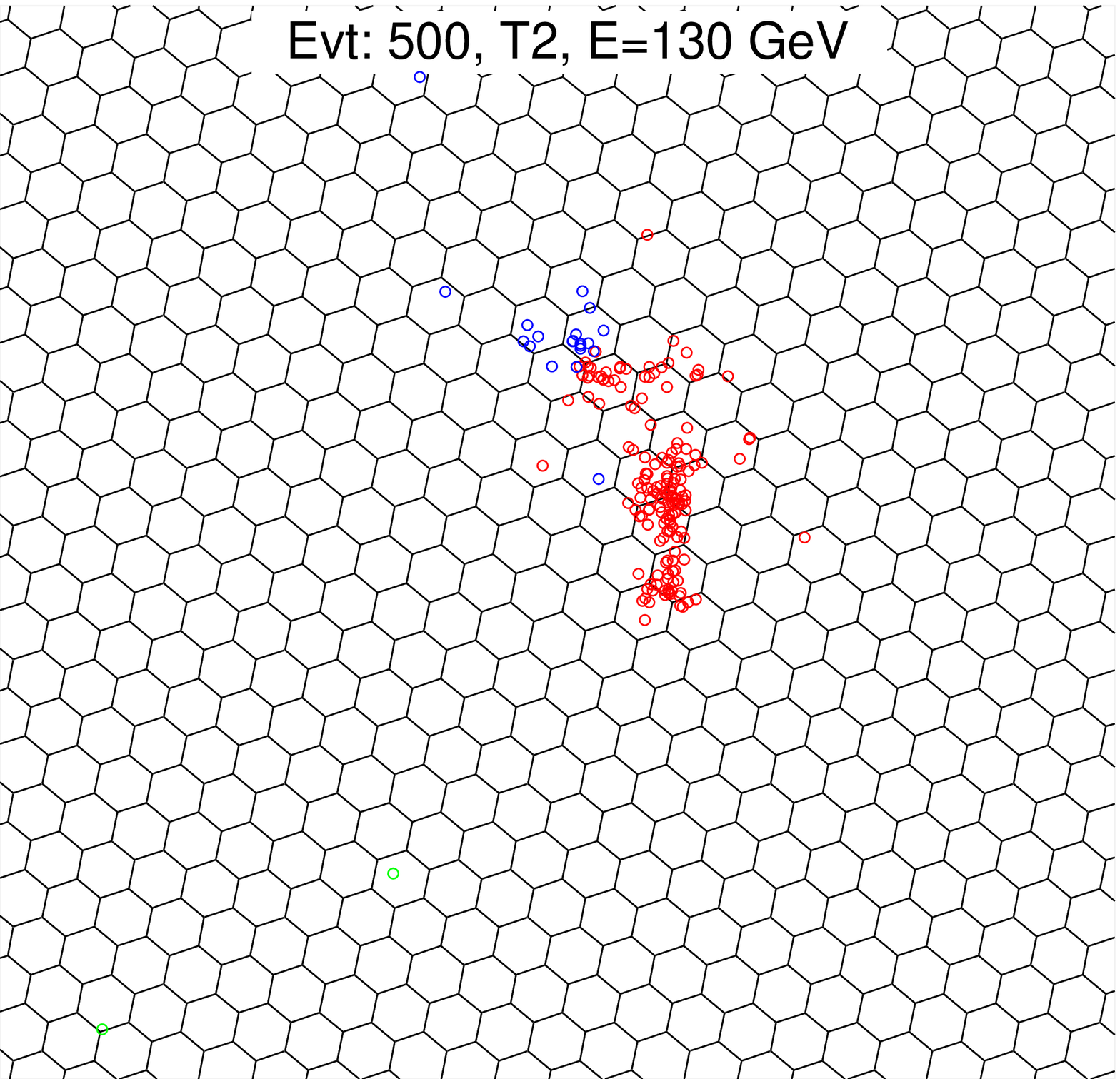}
    \caption{Example image of an event with multiple \ses\ (the top panel) and a \ses - dominated event (bottom panel). 
Each circle represents the position of a single phe in a LST camera coordinates (corresponding to the angular direction of the converted Cherenkov photons).
Different colours represent different \ses . 
Gray points (see the top right part of the top plot) show phe produced by non-\ses\ component (in this case by muons).
For better visibility only a part of the camera containing the shower image is shown.
Black hexagons show individual pixels of the camera. }
    \label{fig:image}
  \end{center}
\end{figure}

In Fig.~\ref{fig:image} we show an example image of a \ses -dominated event and an event composed of multiple \ses .
%
Events composed of multiple \ses\ can show clear, distinct features connected with the development of individual \ses\ through the atmosphere (compare e.g. red and blue points in the top panel of Fig.~\ref{fig:image}), however parts of the image produced by different \ses\ might be also registered at similar angular direction (compare green and red points in the same panel).
On the other hand events with a single dominating \ses\ (see the bottom plot in Fig.~\ref{fig:image}) are more regular and will be able to imitate $\gamma$ rays more effectively.

We perform the image cleaning procedure using the default \progname{chimp} two-pass image cleaning.
The algorithm first searches for pairs of neighboring pixels with $\ge C_1$ phe each (core pixels). 
Pixels with $\ge C_2$ phe and which have a core neighbor are also selected as part of the shower image (boundary pixels). 
The first pass is done with $C_1$-$C_2$ threshold of 6-3 phe and 8-4 phe for LST and MST telescopes respectively.
Next, the time structure of the event (i.e. a linear fit to the signal arrival time vs. position along major image axis) is reconstructed from the pixels surviving the first pass of the cleaning and the signal extraction is redone in all the pixels in a smaller region of interest. 
The extraction of the signal in smaller region of interest allows a second pass of cleaning with lower thresholds (4-2 phe). 
We then calculate the Hillas parameters of each cleaned image \citep{hi85}. 
We exclude from the analysis images of poor quality by applying a cut in minimum size of the image of 50\,phe.
Afterwards stereo parameters are computed from the remaining images. 
The stereo reconstruction is performed using the standard \progname{chimp}-\progname{MARS} chain. 
The direction of the shower is estimated as the point on the camera which minimizes the sum of squares of distances to the main axes of the images.
Next, from each telescope a line in the direction of the image COG is constructed. 
The height of the shower maximum is reconstructed by finding a plane that is minimizing dispersion of crossing points of those lines. 
In both minimization procedures each image is weighted according to its brightness and shape, since these are related to how accurately it matches the shower axis. 

We perform the $\gamma$/hadron separation using multidimensional decision trees, the so-called Random Forest (RF) method \citep{al08}. 
Hillas parameters of a given image (size, width, length, fraction of size in two brightest pixels) together with stereo reconstruction parameters from the whole event (impact and the height of the shower maximum) as well as estimated energy of the event are used to calculate the \emph{Hadronness}$_i$ value of i-th telescope. 
The global \emph{Hadronness} value is computed from averaging individual \emph{Hadronness}$_i$, weighted with the square root of the size (an empirical recipe to give more weight to better-defined shower images). 
The RF method is also used to estimate the energy of each event. 
The weighting of the global estimated energy of the event is done with the inverse square of the uncertainty of energy estimated from Hillas parameters of a given telescope together with the stereo parameters. 
As we are interested in $\gamma$-like background (at a given energy) we train the energy estimation on a subsample of $\gamma$ rays and apply it to the samples of protons, helium and electrons. 
To account for the effect of the GF both the \emph{Hadronness} and estimated energy training are done independently for each of the two simulated azimuths. 
To evaluate the effect of \ses\ and \spis\ on typical observations we calculate G80 cuts, i.e. a cut in \emph{Hadronness} that at a given estimated energy preserves 80\% of $\gamma$ rays. 
In order to investigate background for typical CTA sources we apply a cut in the reconstructed source position of the background events in order for it to lie withing 1.5$^\circ$ from the camera center. 
In such region the angular acceptance is close to constant. 
For computation speed reasons, in order to have also significant statistics at higher energy part of the spectrum all the simulations were done with spectral index $-2$ (see Table~\ref{tab:mc_tel}).
To reproduce the proper energy spectrum of the background components we applied event-wise simulated energy depended weights. 
Unless specified otherwise proton, helium and electron events are reweighted to a power-law with a spectral slope of $-2.73$,  $-2.70$ and $-3.15$ respectively and $\gamma$ rays to a power-law with a spectral slope of $-2.6$. 

\begin{figure}[t]
\centering
\includegraphics[width=0.45\textwidth]{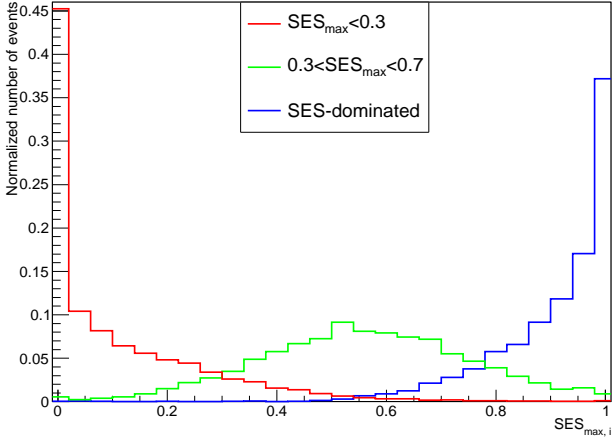}
\caption{Distribution of \ses$_{\max,i}$ for \ses -dominated (\ses$_{\max}>0.7$, blue) proton events compared with proton events without a dominating \ses\ ($0.3<$\ses$_{\max}<0.7$, green and \ses$_{\max}<0.3$, red) for estimated energy range of 12-50\,GeV and LST-subarray.
}\label{fig:onevsall}
\end{figure}

\ses$_{\max}$, being a global parameter calculated from all the triggering telescopes, might in principle hide some information about telescope-wide distribution of \ses s. 
An extreme example would be two \ses s of similar energy produced high in the atmosphere with sufficient angular separation so each of them is seen by a different telescope. 
In order to evaluate if \ses$_{\max}$ is sufficient to classify an event, or if an information from individual telescopes (i.e. \ses$_{\max,i}$, fraction of light produced by the most dominant in that telescope \ses ) is needed, we compare the distributions of \ses$_{\max,i}$ for different \ses$_{\max}$ (see Fig.~\ref{fig:onevsall})
The distribution is mostly concentrated in the ranges defined by binning of \ses$_{\max}$. 
Only a small tail to higher or lower values of \ses$_{\max,i}$ is observed, hence one can conclude that most of the information about the dominating \ses\ is already given by \ses$_{\max}$. 

\section{Results}\label{sec:res}
\begin{figure}[t]
\includegraphics[trim=0 0 0 33, clip, width=0.45\textwidth]{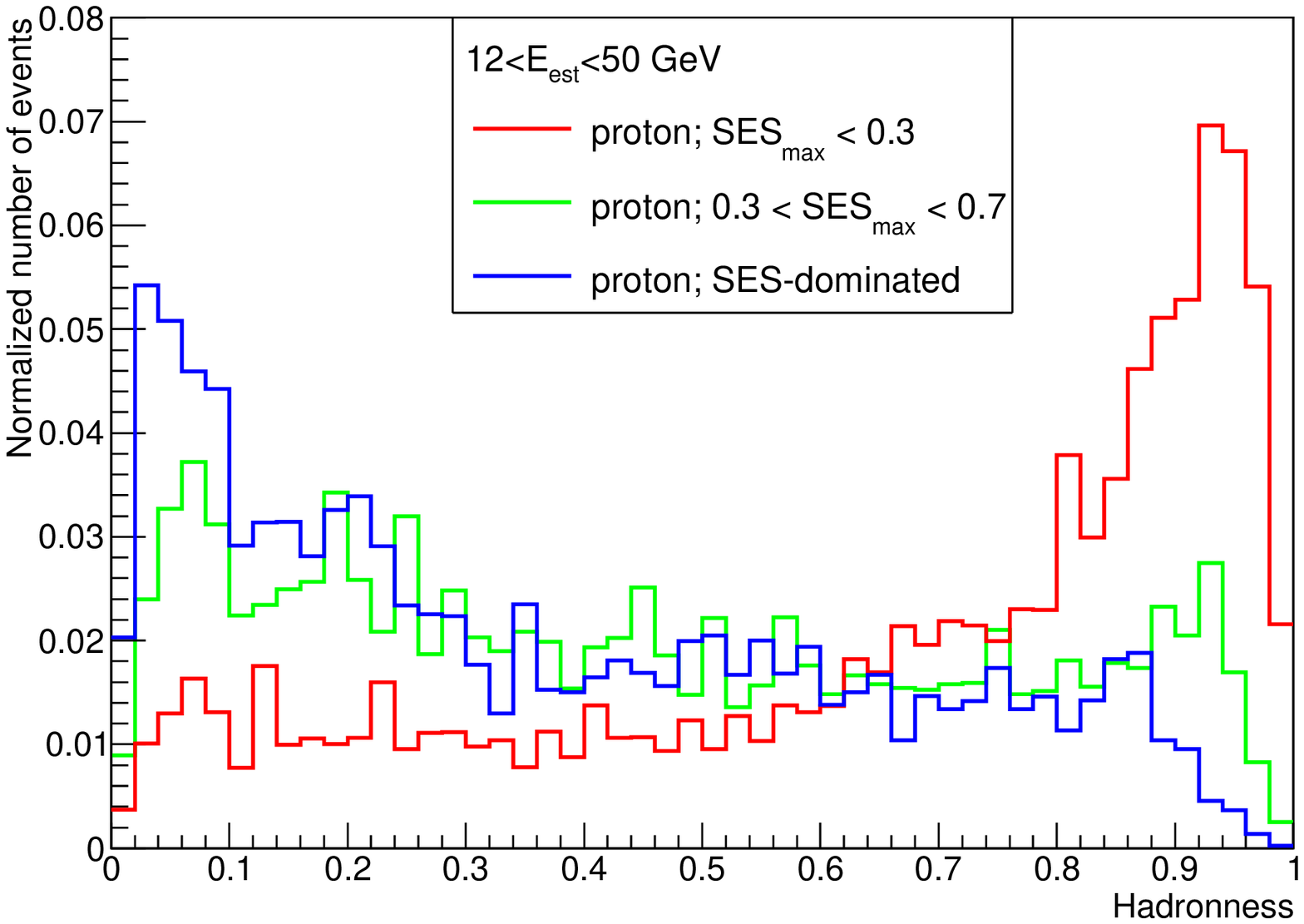}
\includegraphics[trim=0 0 30 33, clip, width=0.45\textwidth]{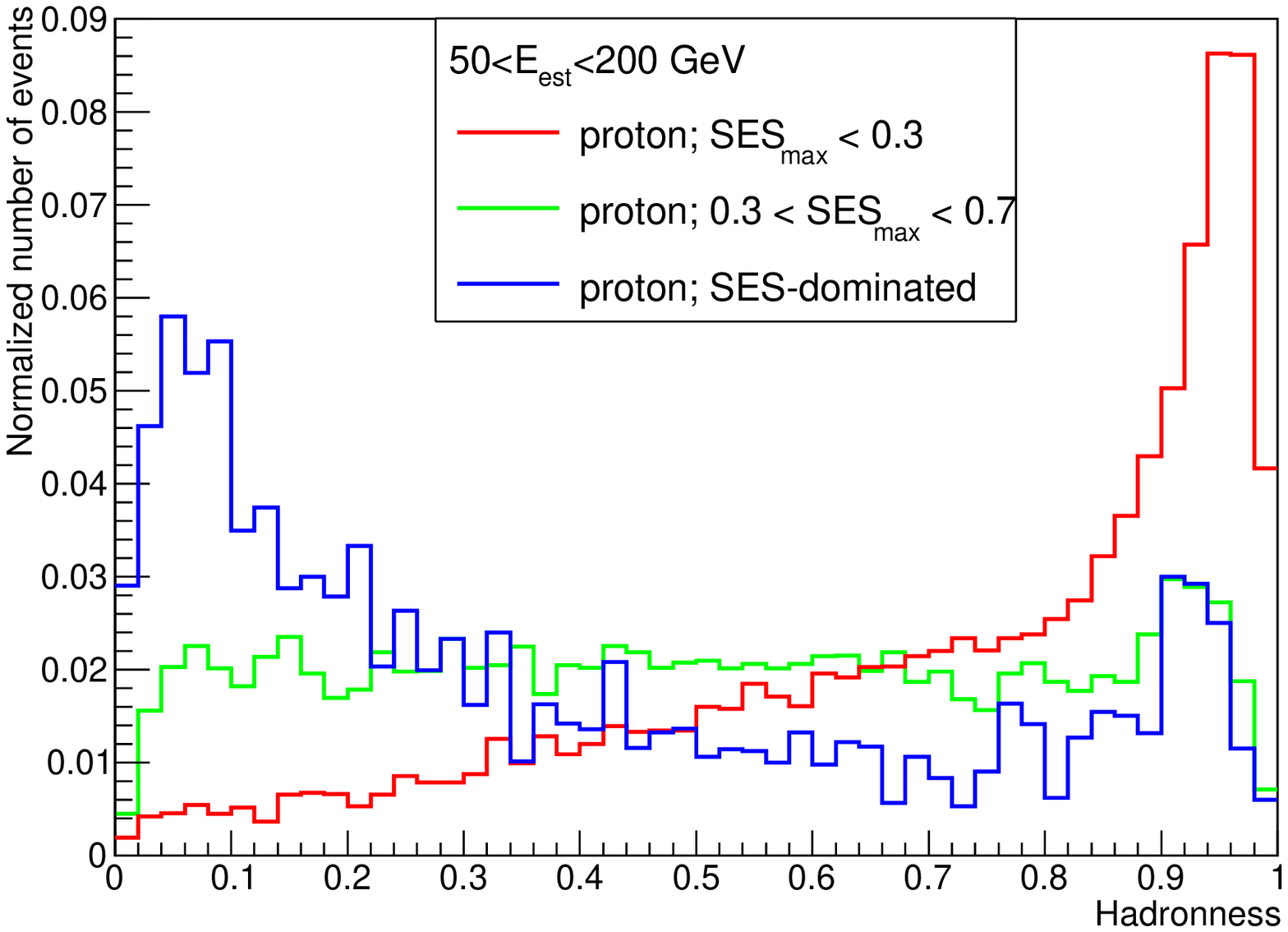}
\caption{Distribution of \emph{Hadronness} for \ses -dominated (\ses$_{\max}>0.7$, blue) proton events compared with proton events without a dominating \ses\ ($0.3<$\ses$_{\max}<0.7$, green and \ses$_{\max}<0.3$, red). 
The top panel shows estimated energy range of 12-50\,GeV for the LST-subarray.
The bottom panel shows estimated energy range of 50-200\,GeV for the full (MST+LST) array.
For each histogram, the sum of bin values is normalized to 1. 
}\label{fig:vshadr}
\end{figure}

As a first step we compute the distribution of the aggregated $\gamma$/hadron separation parameter, \emph{Hadronness}, for events with a different dominance of the largest \ses .
In Fig.~\ref{fig:vshadr} we show such distributions for the lowest energies accessible to the LST sub-array (top panel) and for the energy range from which the MST start to dominate in the full array (bottom panel). 
In both cases there is a clear difference between \ses -dominated and \ses -not-dominated events. 
The former produce a peak at low \emph{Hadronness} and thus efficiently imitate showers initiated by $\gamma$ rays. 
On the other hand, events without a dominating \ses\ are classified with a \emph{Hadronness} value mainly close to 1 and thus are easily rejected from the analysis. 
We have checked that a very similar trend is also observed for events with/without a dominating \spis . 

\begin{figure}[t]
\centering
\includegraphics[trim=0 0 0 35, clip, width=0.45\textwidth]{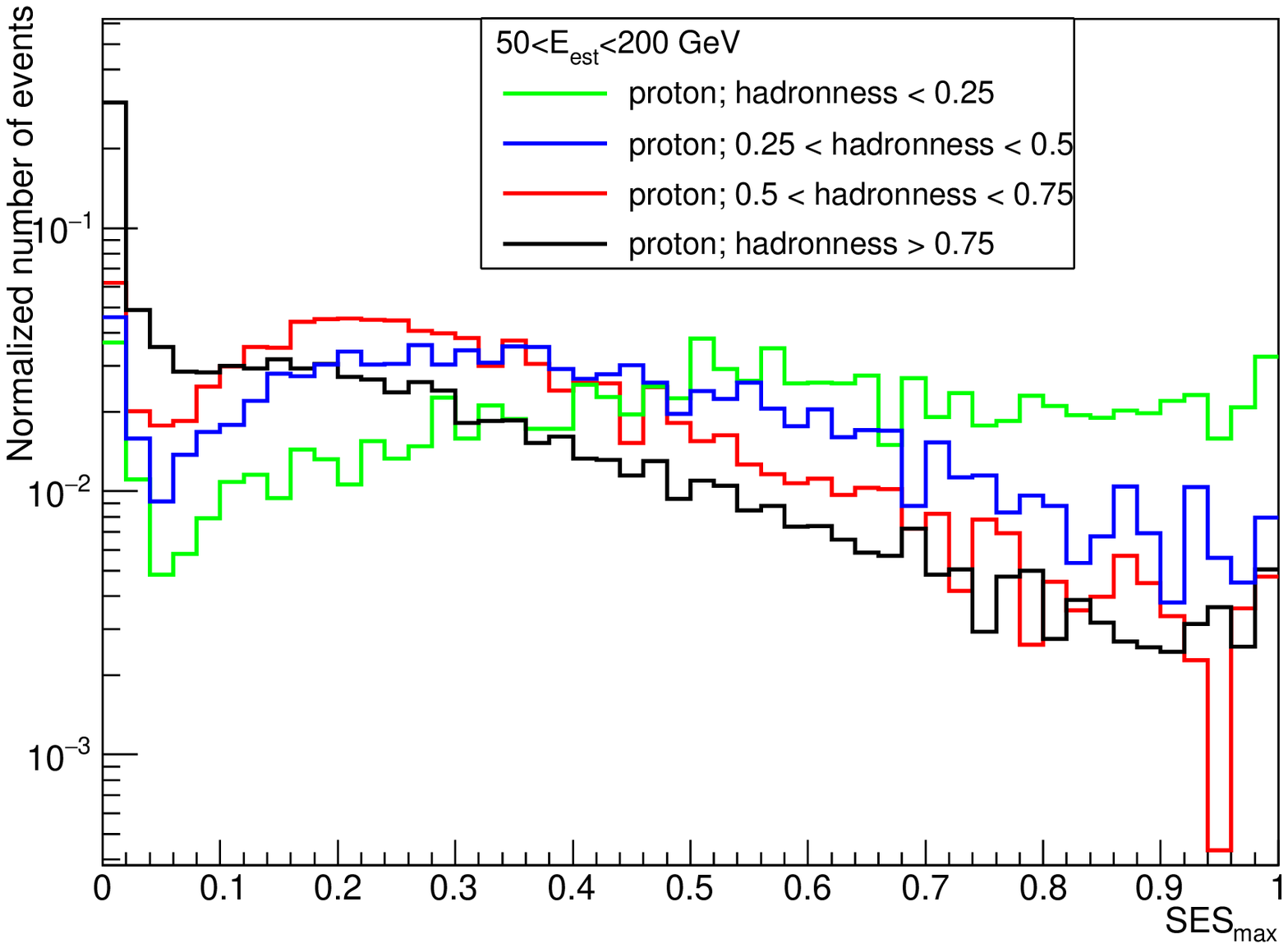}
\includegraphics[trim=0 0 15 35, clip, width=0.45\textwidth]{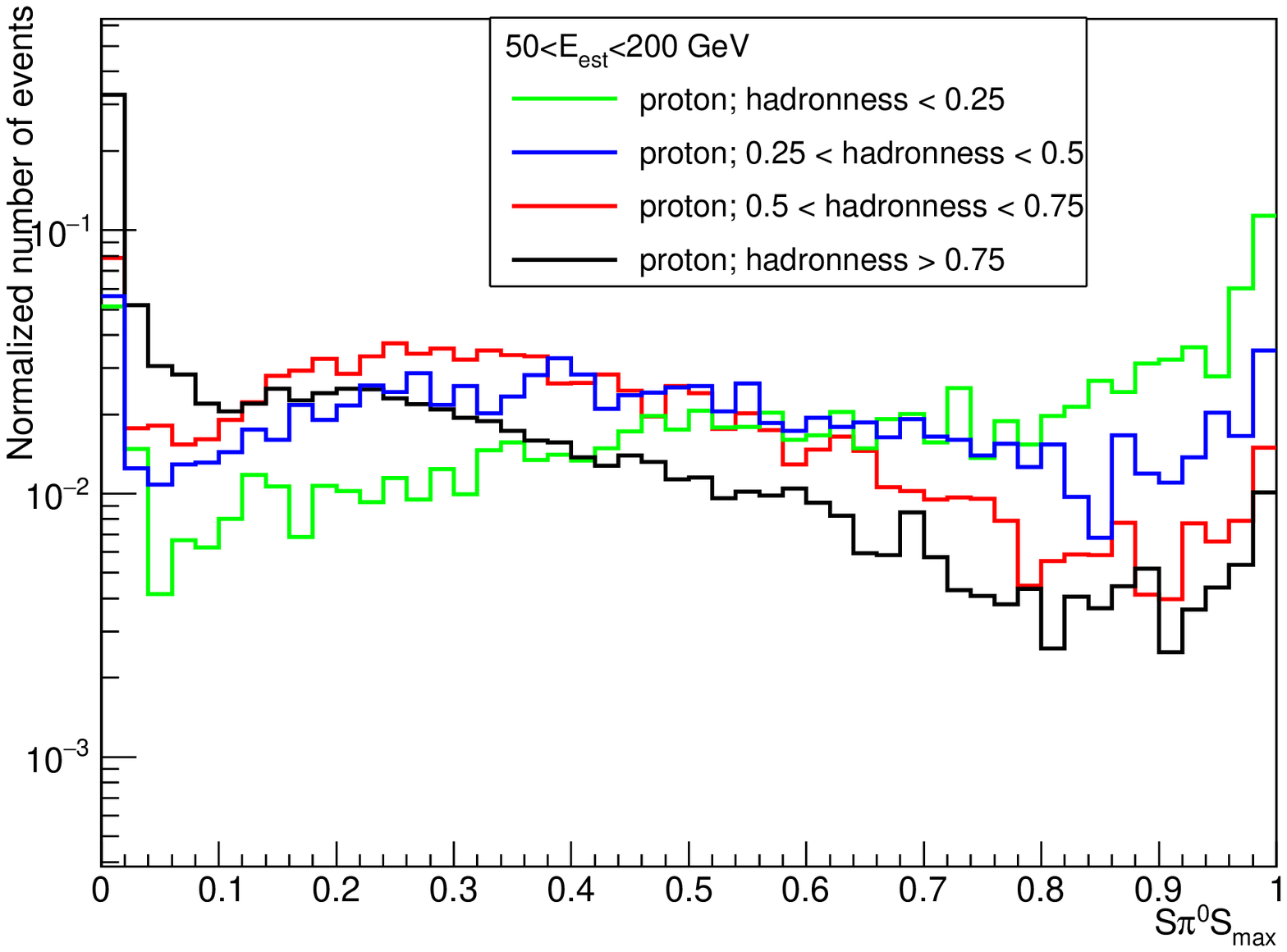}
\caption{Distribution of \ses$_{\max}$ (the top panel) and \spis$_{\max}$ (the bottom panel) for proton events in different bins of \emph{Hadronness} (see legend). Full MST+LST array is used and only events with estimated energy between 50 and 200 GeV are plotted. 
For each histogram, the sum of bin values is normalized to 1.
}\label{fig:vssmax}
\end{figure}

In Fig.~\ref{fig:vssmax} we show the distributions of \ses$_{\max}$ and \spis$_{\max}$ parameters for different bins of \emph{Hadronness} parameter. 
Consistently with what was shown in Fig.~\ref{fig:vshadr} events with low \emph{Hadronness} value have often high \ses$_{\max}$ (and \spis$_{\max}$). 
Comparing the two panels of Fig.~\ref{fig:vssmax}, the \spis$_{\max}\approx1$ peak for low \emph{Hadronness} values is much more pronounced than the corresponding peak at \ses$_{\max}\approx1$. 
Such single \spis\ events are most probably composed of two \ses\ of comparable size.

\begin{figure}[pt]
\includegraphics[trim=0 0 24 25, clip, width=0.45\textwidth]{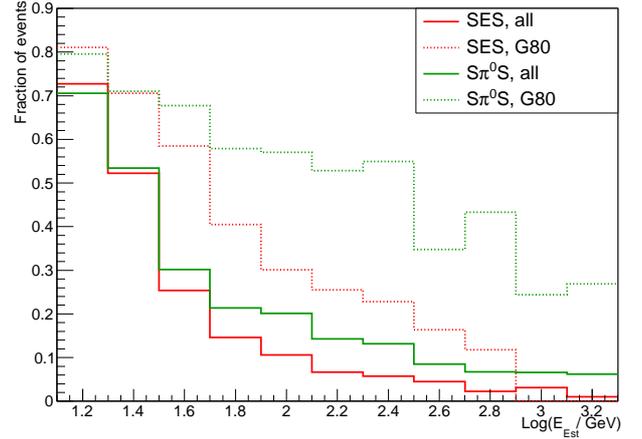}
\caption{Energy dependent fraction of \ses -dominated (red) and \spis -dominated (green) events before a \emph{Hadronness} cut (solid), and after G80 cut (dotted) for the full array. 
}\label{fig:frac}
\end{figure}

In Fig.~\ref{fig:frac} we present the fraction of \ses -dominated  and \spis -dominated events.
The fraction of both \ses - and \spis -dominated events drops down fast with increasing energy, as more individual \ses\ and \spis\ are produced in a shower and can be observed by the telescopes. 
As the showers composed of multiple \ses\ or \spis\ are much easier to separate, after a cut in \emph{Hadronness} the fraction of \ses - and \spis -dominated events is much higher.
At 100\,GeV it reaches 34\% and 57\% respectively.  
In \ref{app:clean} we check if the applied by us cleaning algorithm has any significant impact on the computed fraction of \ses -dominated events. 
No strong influence is found.

\begin{figure}[pt]
\includegraphics[width=0.45\textwidth]{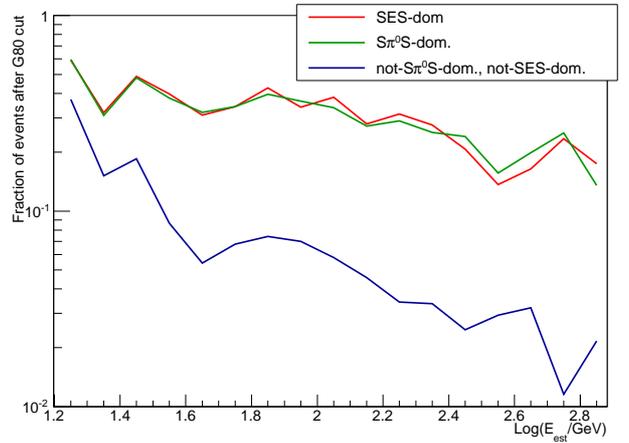}
\caption{Energy dependent fraction of events surviving G80 cut among following groups: \ses -dominated (red), \spis -dominated (green), and non \spis -dominated (blue)}\label{fig:sep}
\end{figure}

In Fig.~\ref{fig:sep} we present the separation power for different classes of events. 
As expected from Fig.~\ref{fig:frac}, both the \ses -dominated and \spis -dominated events are difficult to distinguish from $\gamma$-ray initiated shower. 
Only about 60\% of such events are rejected with a G80 cuts (note that those cuts reject also 20\% of $\gamma$ rays). 
This is nearly an order of magnitude worse than for events without a dominating \spis\ and improves only very slowly with energy.
It is interesting to note that, despite about twice larger fraction of \spis -dominated events than \ses -dominated events, the separation power of both types of events is very similar. 
This suggests that such single-\spis -double-\ses\ events are still similar to a single \ses -dominated events and thus hard to separate from primary $\gamma$ rays.

\subsection{\ses - and \spis -dominated events from protons and helium}
\begin{figure}[pt]
\includegraphics[trim=0 0 24 25, clip, width=0.45\textwidth]{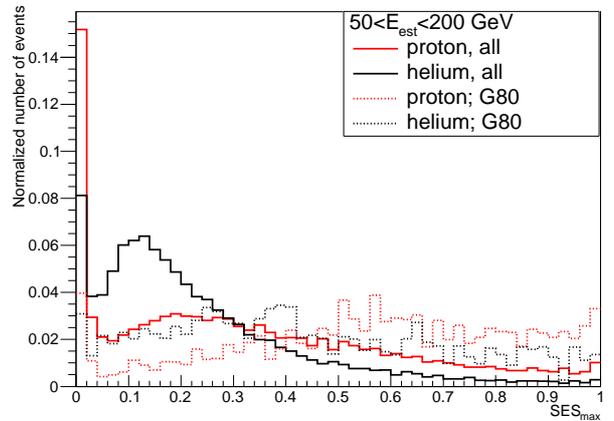}
\caption{Distribution of \ses$_{\max}$ for protons (red) and helium (black), before (solid) and after (dashed) G80 cuts for full array.
Only events with estimated energy in range of 50-200\,GeV are used. }
\label{fig:maxhe}
\end{figure}

In Fig.~\ref{fig:maxhe} we compare the distributions of \ses$_{\max}$ for protons and helium. 
The sharp peak at zero is produced mostly by muon-dominated events. 
The distribution for helium-initiated showers is shifted to lower values. 
It is in line with the typical approximation of helium nuclei as a superposition of four protons with four times smaller energy. 
I.e. in helium a higher number of \ses s is generated and hence a probability for obtaining a single dominant one is smaller.
Interestingly, after \emph{Hadronness} cut, which removes preferentially easier to separate low \ses$_{\max}$ events, the two distributions are more similar. 
The smaller fraction of \ses -dominated events in helium than in protons before the $\gamma$-ray selection seems to be thus one of the reasons why helium-initiated showers are much easier to separate from $\gamma$ rays than proton-initiated ones. 

\begin{figure}[t]
\includegraphics[width=0.45\textwidth]{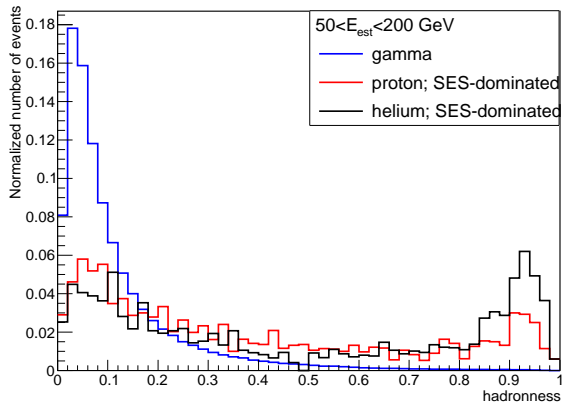}
\caption{Distribution of \emph{Hadronness} for protons (red) and helium (black) \ses -dominated events compared to $\gamma$ rays (blue) for full array.
Only events with estimated energy in range of 50-200\,GeV are used. }
\label{fig:hadrhe}
\end{figure}

\begin{figure}[t]
\includegraphics[width=0.45\textwidth]{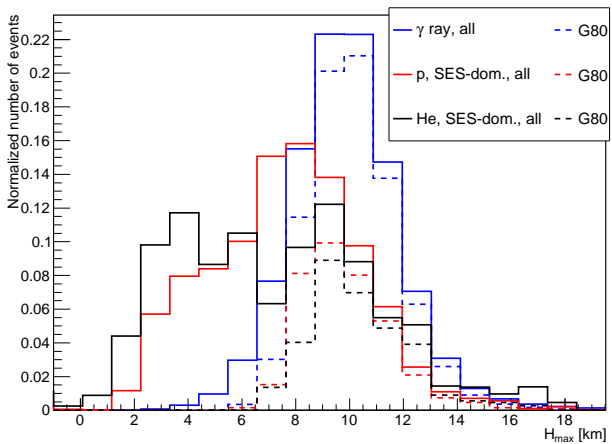}
\caption{Comparison of the distribution of the reconstructed height of the shower maximum (measured above the telescopes) for $\gamma$ rays (blue) and \ses -dominated proton (red) or helium (black) events. 
Solid and dashed lines show respectively the distribution before and after cut in \emph{Hadronness}. 
Full array and events with estimated energy in range of 50-200\,GeV are used. }
\label{fig:hmaxcomp}
\end{figure}

Fig.~\ref{fig:hadrhe} shows the \emph{Hadronness} distribution for \ses -dominated events stemming from protons and helium showers.
\ses -dominated events originating from both proton and helium partially separate from $\gamma$ rays (see the second peak at  \emph{Hadronness}$\,\gtrsim\!0.8$). 
For helium events there is a small preference towards higher hadronness values. 
It is caused by a broader distribution of the reconstructed height of the shower maximum for helium than for protons (see Fig.~\ref{fig:hmaxcomp}).
This might be connected with a higher chance of observing (and classifying as a \ses -dominated event) a helium event composed mostly from a \ses\ produced in the second, or later interaction. 

\subsection{Background composition}
In order to derive the composition of the expected background events for the CTA-North observatory we introduce following classes of events:
\begin{itemize}
\item \ses -dominated events (\ses$_{\max}>0.7$, as before)
\item \spis -dominated, but not \ses -dominated events (\spis$_{\max}>0.7$, \ses$_{\max}<0.7$, i.e. events composed of a dominating $\pi^0$ or $\eta$ subcascade which decays into two (or three) \ses , all observable by the telescopes)
\item muon-dominated, i.e. events with a fraction of observed light produced due to muons above 70\% (by definition those events cannot be \ses - or \spis - dominated)
\item remaining hadronic background events, which are mainly a combination of multiple \ses\ and muons. 
\end{itemize}
We use a combination of proton and helium MCs in a ratio as measured by \cite{ha04}. 
We include also contribution of cosmic ray electrons following \cite{ag14}. 

\begin{figure}[t]
\includegraphics[width=0.49\textwidth]{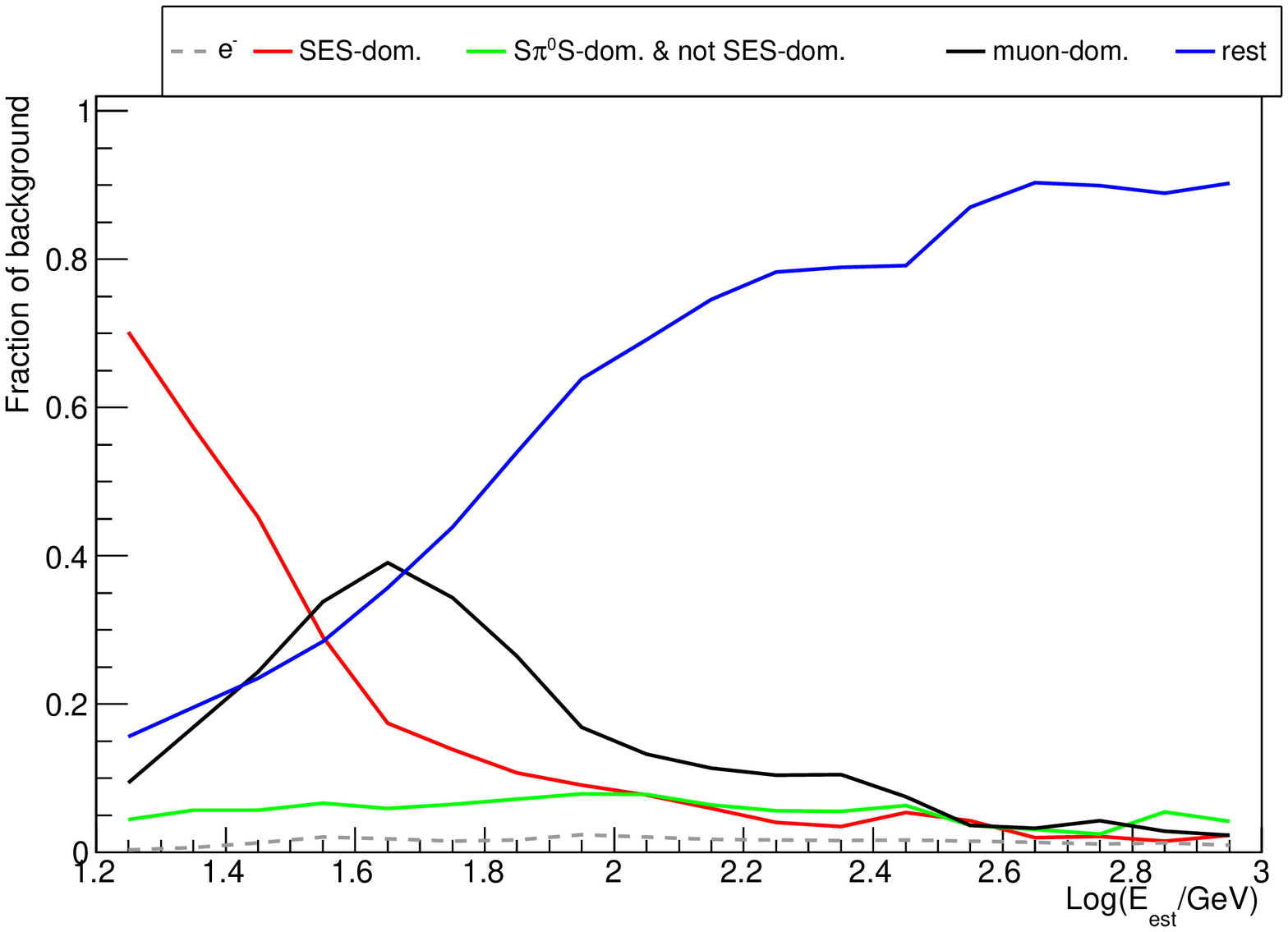}
\includegraphics[width=0.49\textwidth]{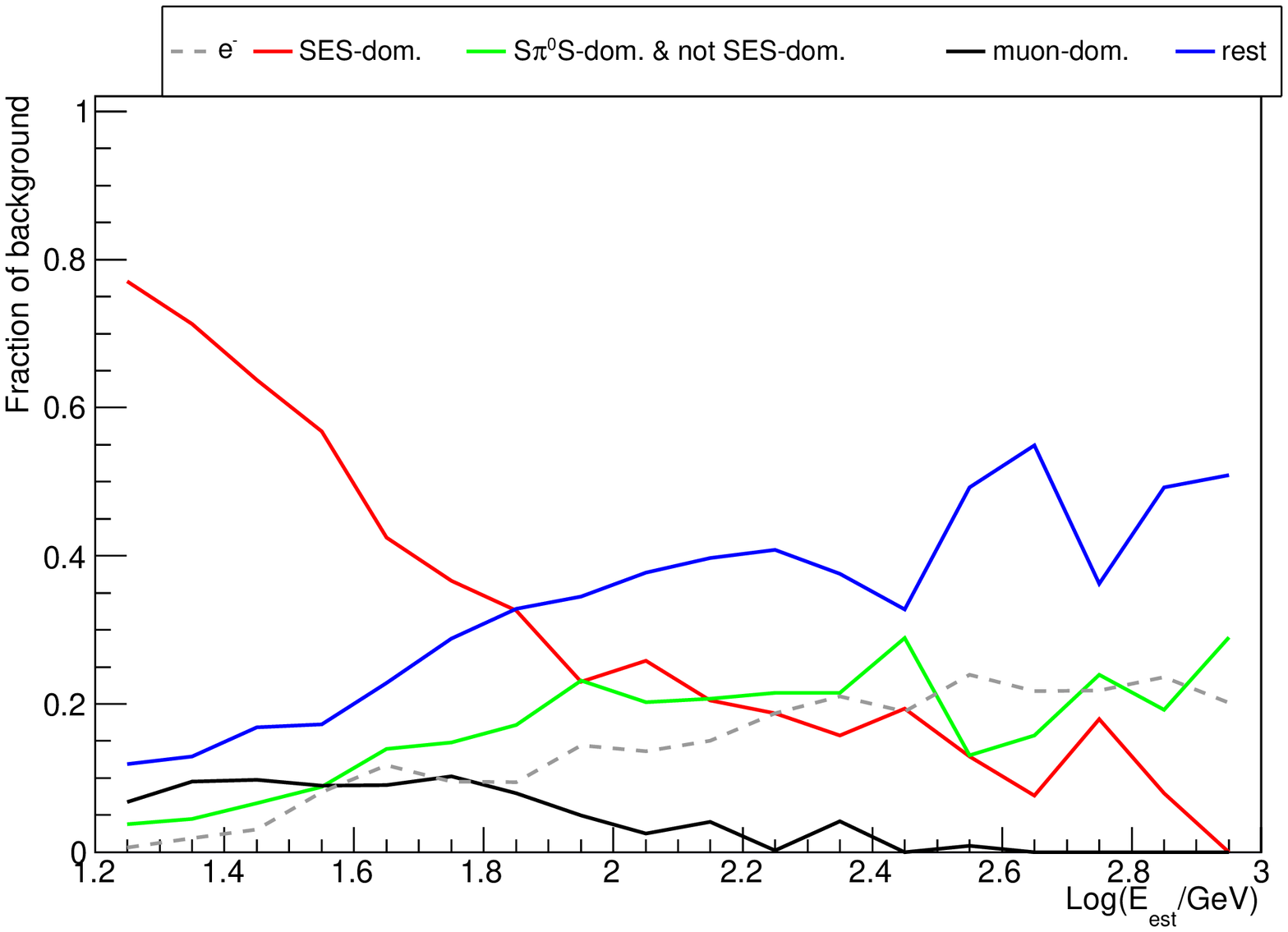}
\caption{Fraction of different classes of background events before (top panel) after (bottom panel) G80 cuts for the full array:
\ses -dominated events (red), 
\spis -dominated but not \ses -dominated (green), 
events with $>70\%$ light registered light induced by $\mu^\pm$ (black),
other (i.e. mixed) hadronic events (blue).
Both proton and helium simulations are used together in proportions as measured by \cite{ha04}.
Fraction of cosmic ray electron background normalized according to \cite{ag14} is shown as gray dashed curve.}\label{fig:comp}
\end{figure}

As can be seen in Fig.~\ref{fig:comp} the background composition is highly dependent on the estimated energy of the shower.
It also changes dramatically after applying $\gamma$/hadron separation cuts due to different separation power of such cuts for different event classes.
At the lowest energies ($\lesssim30$\,GeV) the most important background component is formed by \ses -dominated events. 
As those events separate very badly from $\gamma$ rays, after applying G80 cuts their importance is further enhanced, making them the dominant background up to $\sim$70\,GeV. 
At energies $\sim40$\,GeV there is a large component of muon-dominated background (up to nearly 40\%), which is however easily rejected by G80 cuts, resulting in the final contribution to remaining background at the level of about 10\%. 
\spis -dominated but not \ses -dominated events form only a small fraction (5-10\%) of the background before the cuts, however their poor rejection raises their importance after G80 cuts.
They constitute about 20\% of the remaining background above $100$\,GeV. 
The contribution of mixed events raises fast with the energy. 
Even after the G80 cuts they contribute about 40\% to the remaining background, similar fraction as \ses - and \spis -dominated events together. 
As expected after $\gamma$/hadron separation cuts the contribution of electrons raises with the energy. 
Above 100\,GeV it reaches a similar fraction as \ses -dominated events. 

\subsection{Rejection of \ses -dominated events}
A \ses -dominated event is an electromagnetic cascade and hence it is expected that they are difficult to separate from $\gamma$ ray initiated showers. 
For example the commonly used in IACT analysis Hillas parameters have very similar distributions for \ses -dominated events and $\gamma$ rays \cite{sob2007,so15a}.
Let us consider a simple case of a CR proton interacting with an air nucleus producing at depth $t_1$ a $\pi^0$ particle decaying into two $\gamma$ rays, with energies of $\sim$100\,GeV and $\ll100$\,GeV respectively. 
The first $\gamma$ ray will interact with another air nucleus at depth $t_2$ and initiate an electromagnetic cascade. 
If the telescopes register light only from this sub-cascade, the image will be indistinguishable from a primary $\gamma$ ray which due to fluctuations had its first interaction at the same depth of $t_2$. 
As the first interaction depth cannot be measured directly, in such a case the only parameter which can be used to reject those events is related to the height of the shower maximum. 
The average depth of the first interaction for a 100\,GeV proton is about $90\,\mathrm{g\,cm^{-2}}$ (see e.g. \citealp{mi94}), while the fluctuation of the shower maximum for a 100\,GeV $\gamma$ ray is about $60\,\mathrm{g\,cm^{-2}}$ (dominated by the fluctuations of the first interaction).
Hence, some separation of \ses -dominated events from $\gamma$ rays is possible on the base of the reconstructed height of the shower maximum, however with a large overlap of the two distributions. 
In fact such a parameter is already used in IACT analysis as it is also a powerful muon rejection tool (see e.g. \citealp{al12}).

We want to test if the currently achieved in CTA simulations rejection power of \ses -dominated events is already limited by physics of the showers, or can it be still improved by using e.g. better estimation of the height of the shower maximum. 
To evaluate how strong suppression of \ses -dominated events with a cut in the height of the shower maximum is possible we first perform a toy MC study. 
Next we compare it with the results obtained with the full simulations.  
In the toy MC study for a given energy of $\gamma$ ray, $E$, we simulate 1000 showers using \progname{CORSIKA}.
The longitudinal distribution of each shower is fitted with a Gaisser-Hillas profile \citep{gh77,he98} used in \progname{CORSIKA}. 
We extract from the fit the depth of the shower maximum and construct a distribution of it, hereafter $D_\gamma$. 
In the toy MCs we assume that a \ses -dominated event that can mimic a $\gamma$ ray of energy $E$ must have also a similar energy to $E$ (note that most of the primary proton energy should go into a single subcascade if no other subcascade or muon is observed). 
We assume that the \ses\ starts at the depth of the first interaction of protons, following the proton-air cross section of \cite{mi94}.
As the energy dependence of the cross section is only logarithmic, the assumption about similar energies of $\gamma$ rays and protons should not affect the results strongly.
The depth of the shower maximum for a \ses\ will then be a sum of the depth of the first interaction (drawn from an exponential function) and the depth of the shower maximum for a $\gamma$ ray (drawn from $D_\gamma$). 
We then construct a distribution of such computed depth of the shower maximum of \ses\ events and compare it with $D_\gamma$. 
We calculate a cut value in the depth (or equivalently height) of the shower maximum that maximizes the so called Q-factor, i.e. fraction of surviving $\gamma$ rays divided by the square root of the remaining \ses\ events. 
The value of the Q-factor can be understood as the improvement of the sensitivity that such a cut can give if this type of events is the dominating one. 

\begin{figure*}[t!]
\centering
\includegraphics[width=0.45\textwidth]{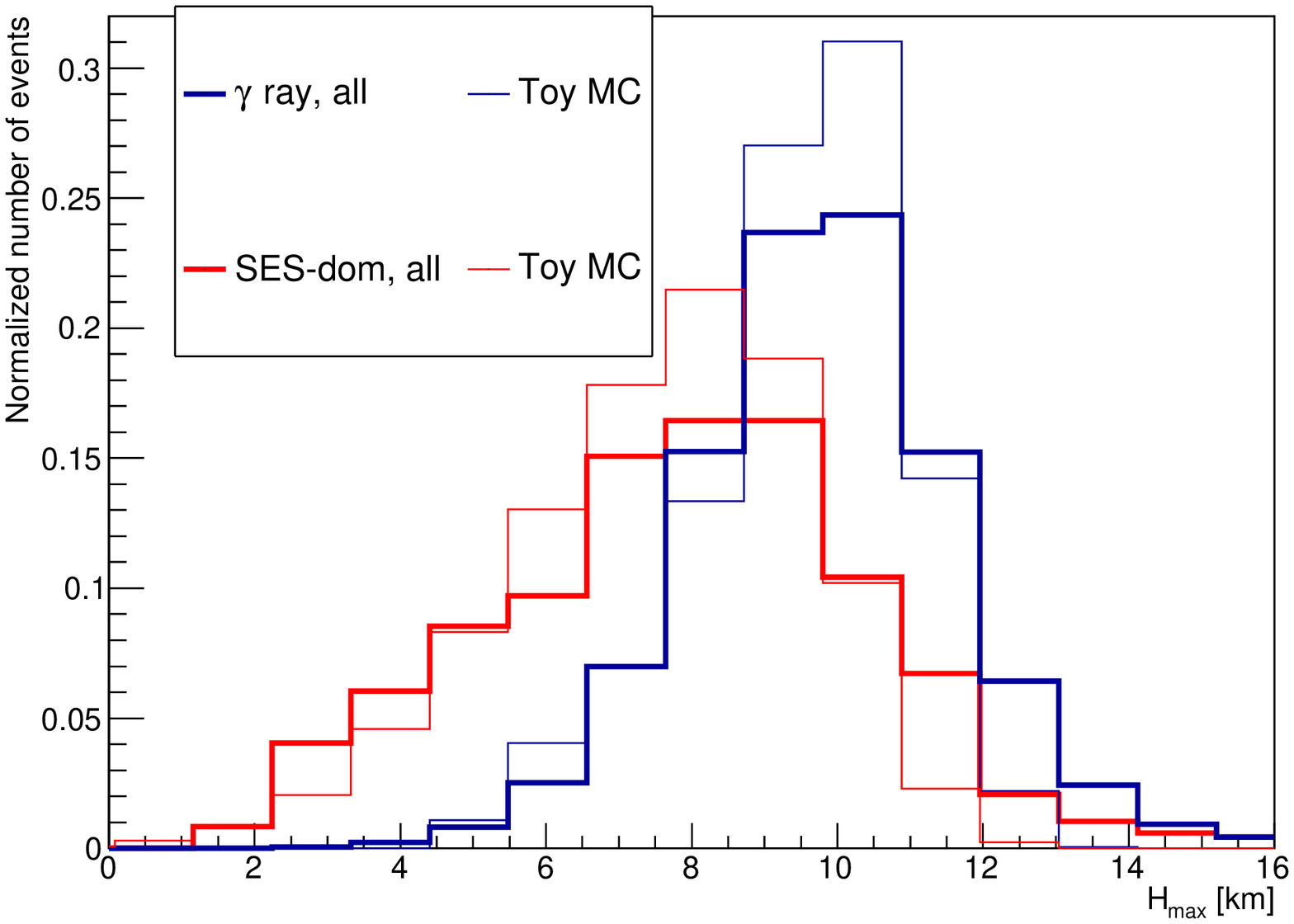}
\includegraphics[width=0.45\textwidth]{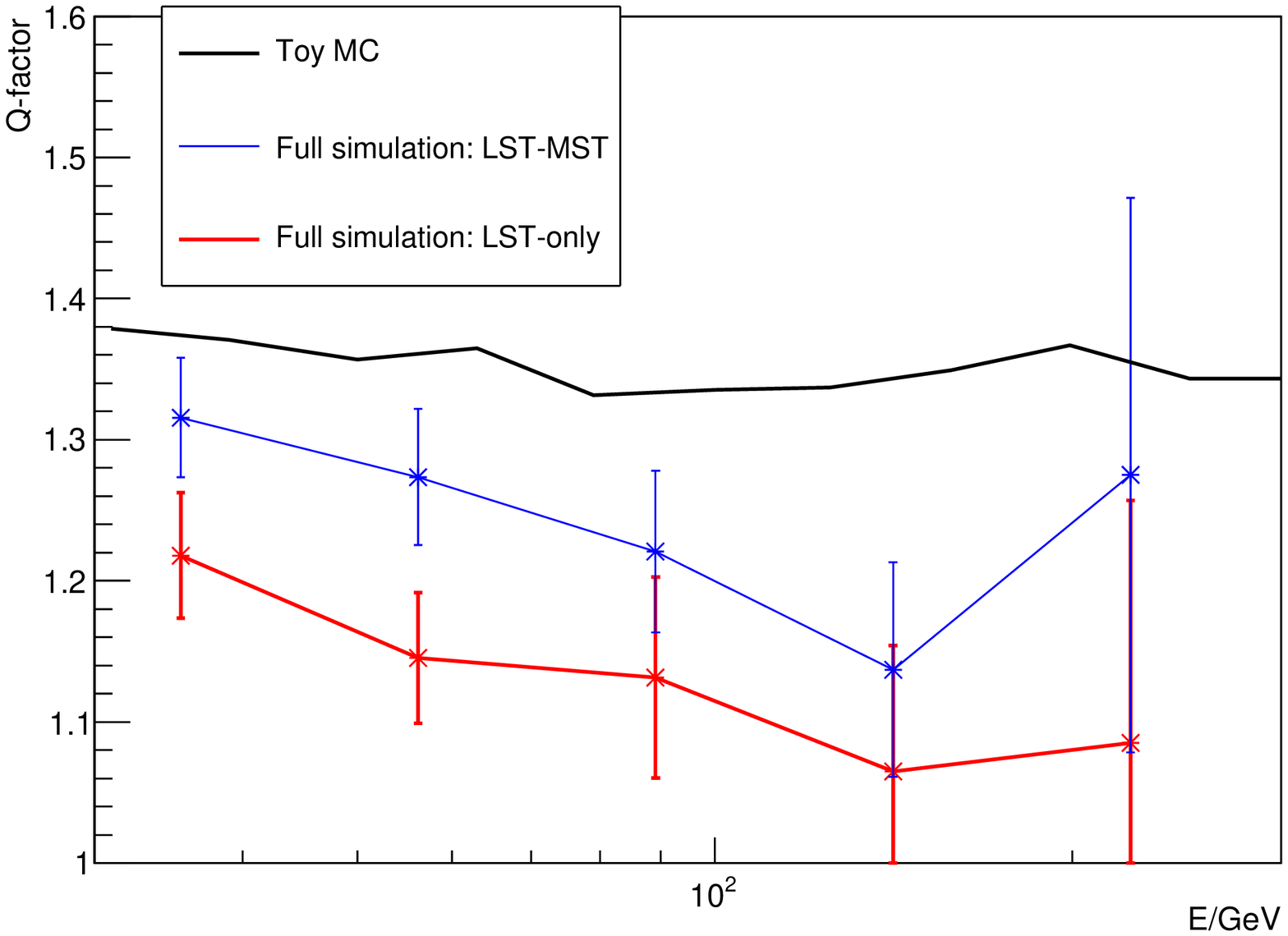}
\caption{
Left panel: the comparison of the distribution of the reconstructed height of the shower maximum for $\gamma$ rays (blue) and \ses -dominated proton events (red) obtained in the full simulations (thick lines) and toy MC (thin lines)
Full LST-MST array is used and events with reconstructed energy in range 70-140\,GeV are selected. 
Right panel: the best cut quality factor of height of the shower maximum is plotted as the function of the energy.}\label{fig:hmax}
\end{figure*}

The distribution of the corresponding height of the shower maximum and the energy dependence of the Q-factor obtained from the toy MC are compared to the full simulations in the middle panel of Fig.~\ref{fig:hmax}. 
For the full simulation case the distributions of the reconstructed height of the shower maximum for gamma-rays and \ses -dominated events are done in bins of estimated energy. 
The toy model is rather simple and does not take into account a few important effects. 
In particular, due to the atmospheric absorption, the height of the shower maximum observed in Cherenkov radiation shifts to lower heights \citep{so09a}.
In addition, \ses -dominated events that fluctuate deeper into the atmosphere will have an enhanced chance of detection, and if so, their energy is reconstructed as that of an even deeper developing $\gamma$ ray; such effect will be most important close to the energy threshold of the telescopes. 
Nevertheless, at least for energies around $100$\,GeV the toy MCs seem to describe relatively well the distribution of height of the shower maximum obtained from the full simulations.  
The expected Q-factor of the height of the shower maximum cut from toy MC is about 1.35 (see the right panel of Fig.~\ref{fig:hmax}). 
The obtained Q-factor from the full CTA-North array (i.e. LST and MST) simulation is $\sim$6\% worse, while the Q-factor from the LST-subarray is $\sim$14\% worse. 
Somewhat higher values obtained in Toy MC as well as the better performance of the full array with respect to the LST subarray is probably caused by a better stereo reconstruction due to larger multiplicity of telescopes observing a given event. 
It might be however also enhanced by the event selection bias of smaller MST telescopes. 
Events starting deeper into the atmosphere would preferentially trigger these telescopes.
Such events are then easier to reject by a cut in the height of the shower maximum. 
That bias can be responsible for the hint of a higher Q-factor at the lowest energies, which is not reproduced in toy MC. 

Comparing the distribution of the height of the shower maximum before and after G80 cuts (left panel of Fig.~\ref{fig:hmax}) it is clear that the RF is already efficiently exploiting the information of the height of the shower maximum. 

In principle, the separation of \ses -dominated events from $\gamma$ rays might be also based on a search for signatures of direct Cherenkov radiation from the primary particle. 
Most of the \ses -dominated events are produced by protons with energies of the order of 100\,GeV. 
Hence, in the upper parts of the atmosphere their energy will be still below the threshold for the Cherenkov radiation. 
Even if a proton is above such threshold, contrary to higher Z elements, its direct Cherenkov radiation is very weak.
Our simulations show that in more than a half of \ses -dominated events not even a single phe is produced in any of the telescopes from a Cherenkov photon produced by a proton. 
Hence we conclude that any possible separation based on the search of direct Cherenkov light with CTA will be very inefficient. 

\section{Discussion and conclusions}\label{sec:con}
Using standard CTA simulation software and state-of-the-art Cherenkov telescopes analysis methods we have studied different classes of low energy background events. 
In particular we have investigated events composed mainly by a single electromagnetic subcascade, or a pair of electromagnetic subcascades from a decay of a neutral particle. 
We performed full MC simulations of $\gamma$ rays, protons, helium and electrons for one of the most promising arrays designed for CTA-North. 
As expected, \ses -dominated and \spis -dominated are very similar to $\gamma$-ray induced showers, and hence difficult to reject.
Comparing proton and helium simulations, at a given energy the latter have a larger number of smaller \ses , making it rarer to have one dominating \ses . 
This can explain why helium and higher elements, while relatively abundant in the observed CR spectrum, have a rather small effect on the remaining background for Cherenkov telescopes.

After $\gamma$/hadron separation cuts \ses - and \spis -dominated events  constitute $\gtrsim$50\% of the residual cosmic ray background mimicking $\gamma$ rays with energies $\lesssim 100$\,GeV. 
A \ses -dominated event is formed by an electromagnetic cascade in the atmosphere, that is virtually indistinguishable from a $\gamma$ ray if starting at the same depth in the atmosphere.
As the \ses -dominated events on average start $\sim90\mathrm{g\,cm^{-2}}$ deeper, some separation is possible, and already being done, based on the height of the shower maximum.  
By performing comparisons between the full MC simulations and a toy MC we have shown that the currently achieved rejection of \ses -dominated events stemming from the height of the shower maximum estimation is close to the expected natural limit. 
Therefore, no big improvement in the background rejection at the lowest $\lesssim 50$\,GeV energies is to be expected by using more elaborate analysis methods.  
On the other hand, at slightly higher energies of $\sim 100$\,GeV, the fraction of \ses -dominated events cross (at the level of $\sim$20\%) the fraction of events that are \spis -dominated, but not \ses -dominated. 
\spis -dominated events composed of two \ses\ of a comparable size, contrary to \ses -dominated events, are qualitatively different from $\gamma$ rays, however it is intriguing that according to our studies their rejection power is still similar. 
Hence, it is possible that with a rejection method more focused on this type of background higher performance of $\gamma$/hadron separation might be achieved. 
One should note however that at those energies an even larger fraction of the background is produced by mixed events and a non-negligible amount by cosmic ray electrons. 

\appendix 

\section*{Acknowledgements}
This work is supported by the grant through the Polish Narodowe Centrum Nauki No. 2015/19/D/ST9/00616.
DS is supported by the National Science Centre grant No. UMO-2016/22/M/ST9/00583. 
AM acknowledges the support of the ​MultiDark​ ​CSD2009-00064 project of the Spanish Consolider-Ingenio 2010 programme.
We would like to thank CTA Consortium and MAGIC Collaboration for allowing us to use their software and Konrad Bernl\"ohr for helpful discussions. 
We would also like to thank the anonymous journal reviewer for his helpful comments.

This paper has gone through internal review by the CTA Consortium.

\section{Dependence on cleaning algorithm}\label{app:clean}
While the classification of events as \ses - or \spis -dominated that we apply is cleaning independent, the rest of the analysis chain will clearly depend on it. 
For example let us consider an event composed of two \ses , which are separated on the camera.
If the cleaning algorithm excludes for some reason (e.g. lower photon density, or separation in arrival time) the pixels with the signal from one of those \ses , the resulting event would behave during the reconstruction like a \ses -dominated one, despite the fact that it does not have to be classified as such. 
In particular the effect of the double-pass cleaning with signal re-extraction explained in Section~\ref{sec:anal} can be quite complicated.
On one hand, the re-extraction of signal in time bins determined from the core of the image allows us to include also lower intensity pixels into image, without increasing the influence of the NSB. 
This can reveal in a hadronic background events parts of the image produced by a separate \ses , making the event less $\gamma$-like. 
On the other hand the re-extraction of signal could also clean away parts of the image produced by one of the \ses , if they are separated in time/distance space from the rest of the image. 
To investigate if those effects are important for the composition of the $\gamma$-like background we performed a simplified study using the LST subarray and only one Azimuth angle. 
We processed this subsample with only the first stage of cleaning and compared the obtained results with the double-pass cleaning analysis used in the rest of the paper.
Due to the different cleaning, the $\gamma$/hadron separation and energy estimation are trained separately for the 1-pass cleaning sample. 
For the fair comparison of both cleanings we use G80 cuts calculated separately for each of them. 

\begin{figure}[t]
\includegraphics[trim=0 0 0 0, clip, width=0.45\textwidth]{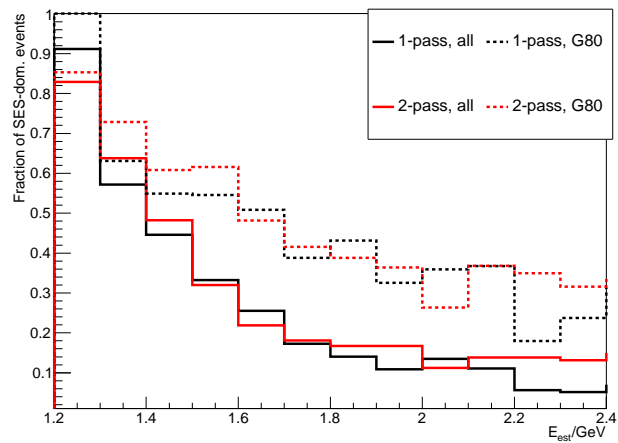}
\caption{
Fraction of \ses - dominated events in proton background for LST subarray and Azimuth angle of $180^\circ$ for single-pass (magenta) and double-pass (cyan) cleaning as a function of estimated energy of the event. 
Solid lines show all the events, dotted lines are event surviving the G80 cuts. 
}
\label{fig:clean}
\end{figure}

In Fig.~\ref{fig:clean} we show the dependence of the fraction of \ses - dominated events on the estimated energy for single-pass and double-pass analysis. 
The fraction of \ses -dominated events after the G80 cuts is similar for the two cleaning approaches. 
Thus, the cleaning method should not have a strong impact on the presented results.



\end{document}